# Crisis and catharsis in atomic physics

Precise measurement of an atomic hydrogen transition resolves the proton size puzzle

Wim Ubachs

Vrije Universiteit, Amsterdam, Netherlands.
Email: w.m.g.ubachs@vu.nl

The spectrum of the simplest atomic species, the hydrogen (H) atom with only a single electron, was initially described with Bohr's theory of 1913 and was refined with newer theories, from Schrödinger's quantum mechanics to Dirac's relativistic formalism and ultimately Feynman-Schwinger-Tomonaga's quantum electrodynamics (QED). The latter includes the effects of virtual particles that emerge from the vacuum and deals with the problem that the energy of a charged point particle is infinite. The comparison between theory and precise measurements of the H atom ran into a crisis in 2010, when measurements on muonic hydrogen (where muons replace electrons) (*1*) led to two different values of the size of the proton, $r_p$ (see the figure). A decade-long period of the "proton size puzzle" spurred renewed experimental activity and many far-reaching hypothetical theories. On page 1061, Grinin *et al.* (*2*) report the precision measurement of the 1S-3S transition to help finally resolve this crisis.

The quantum-level structure of the hydrogen atom can be described for each state as the energy $E$ that depends on quantum numbers $n$, $l$, and $j$ and is the sum of three terms:

$$E(n,l,j) = E_{Bohr}\, f(m_p, m_e) + E_{NS}(r_p, n, l) + E_{QED}(n, l, j) \qquad (1)$$

where $E_{Bohr}$ represents the Bohr structure, $f(m_p, m_e)$ is a small correction involving the proton mass $m_p$ and the electron mass $m_e$, $E_{NS}$ describes nuclear size effects, and $E_{QED}$ represents the QED corrections. $E_{Bohr}$ is proportional to the Rydberg constant $R_\infty$. $E_{NS}$ accounts for an electron penetrating the inner region of the extended proton, where it experiences an attractive force that deviates from the usual Coulomb force.

The entire level structure of the hydrogen atom can be cast by the two unknowns: $R_\infty$, representing the energy scale of all atomic physics and of chemistry, and $r_p$. These two unknowns can be determined from several sets of two precise measurements. The measurement of the 1S-2S interval (*3*) stands out as the most precise (reaching 15-digit accuracy) because the upper 2S level exists for ~1 s and is not affected by the Heisenberg uncertainty principle. This result may then be combined with a second result. Various groups (*4–6*) have measured transitions from the long-lived 2S level to levels with $n$S, $n$P, or $n$D states (called Rydberg states), and these measurements were considered statistically independent and averaged to obtain a set of values for $R_\infty$ and $r_p$. Another approach is in the measurement of the Lamb shift (the 2S-2P splitting) initially performed by Lamb and Retherford in 1948 and improved thereafter (*7*).

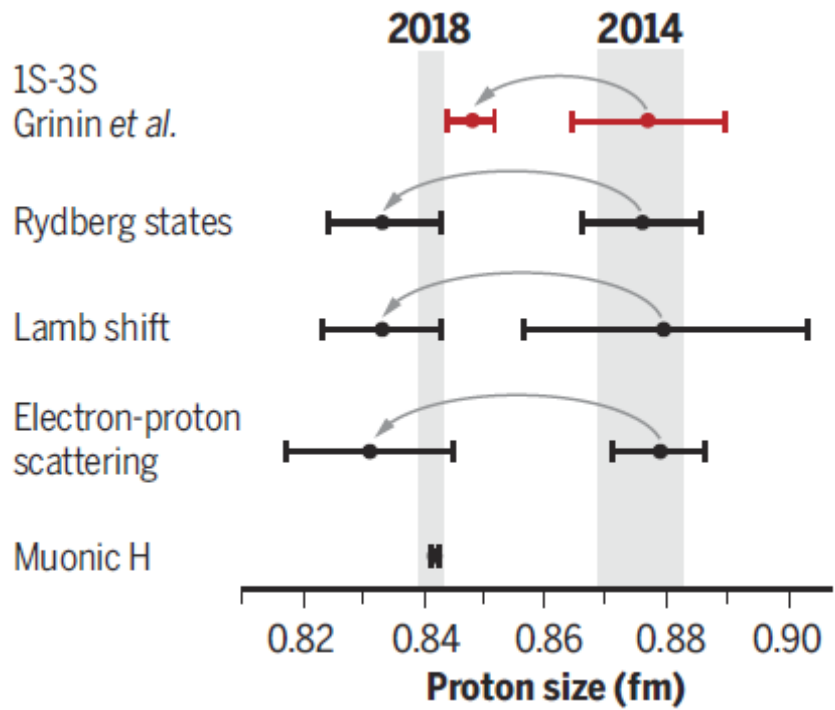

Figure: The right side shows the consistent picture of the proton size of 2010 before the muonic hydrogen measurement. The left side displays the current consistent picture, and arrows indicate the historical correction. Gray zones are the recommended values from the Committee on Data for Science and Technology (CODATA) for 2014 and 2018.

Electron-proton scattering measurements yield a fully independent determination of the proton size (8), although this nuclear physics approach requires a complicated extrapolation to a zero-momentum scattering vector. By 2010, these approaches led to a consistent set of values for $R_\infty$ and $r_p$. In the figure, the values for $r_p$ are plotted, and $r_p$ was determined at 0.88 fm.

Because the Standard Model of physics assumes that the muon exhibits the same physics as the electron (lepton universality), except for being 207 times more massive and prone to radioactive decay, the level structure of muonic hydrogen (μH) should be calculable with Eq. (1). The heavier mass makes the muon overlap more with the nucleus, a reason why comparison between experiment and theory led to a very accurate value for $r_p$ at 0.84 fm (1).

Thus, the atomic physics community was shaken to have the "consistent value" of $r_p$ disagree with this μH measurement well beyond their error limits. This situation spurred enormous experimental activity on the precision spectroscopy of the H atom, as well as tests of the far-reaching hypothesis of the breakdown of lepton universality, and in turn the Standard Model of physics. Painstaking experiments remeasured the Lamb shift (9), excitation to a 4P Rydberg state (10), and electron-proton scattering (11), and all produced $r_p$ of 0.84 fm.

However, in 2018, a continuous-wave laser experiment on the 1S-3S transition (12) still obtained the larger value. The situation is now finally resolved with the more accurate direct frequency-comb laser experiment performed by Grinin et al. on the same 1S-3S transition. The smaller $r_p$ value supports a QED frameworkof electronic and muonic hydrogen described by the two constants, $R_\infty$ and $r_p$ = 0.84 fm. This struggle for consistency in atomic physics should provide an intriguing topic for historians and sociologists of science.

The H atom, and the interesting alternative of the He$^+$ ion, has only a single long-lived excited state, which hampers precision experiments on a set of quantum levels. Molecules such as the $H_2^+$ and $HD^+$ ions support many ro-vibrationally excited states that can all be subjected to precision measurements. Studies of these molecules have already led to a determination of the proton-electron mass ratio (*13, 14*). The neutral $H_2$ molecule, which supports >300 ro-vibrational states with lifetimes of 1 week, could also be used in precision experiments for testing fundamental physics, possibly at 20-digit accuracy (*15*).